\documentclass{PoS}
\usepackage{graphicx}
\usepackage{float}
\usepackage{lineno}
\graphicspath{{/home/vikas/Dropbox/repos/icrc_2017_proc/plots/}}
\title{HAWC High Energy Upgrade with a Sparse Outrigger Array}

\ShortTitle{HAWC High Energy Upgrade with a Sparse Outrigger Array	}

\author{\speaker{Vikas Joshi}$^a$,{Armelle Jardin-Blicq}$^a$ for the HAWC Collaboration$^b$
\\$^a$Max Planck Institut f\"{u}r Kernphysik, Heidelberg, Germany\\$^b$For a complete author list, see \textcolor{blue}{ http://www.hawc-observatory.org/collaboration/icrc2017.php} \\E-mail: \email{vikas.joshi@mpi-hd.mpg.de}
\\  \hspace{11mm}      \email{armelle.jardin-blicq@mpi-hd.mpg.de}}

\abstract{The High Altitude Water Cherenkov (HAWC) gamma-ray observatory consists of 300 water Cherenkov detectors and has been fully operational since March 2015 in central Mexico. It detects cosmic- and gamma-ray showers in the TeV energy range. For multi-TeV energies, the shower reconstruction and hence the performance of the detector is affected by the partial containment of the showers within the array. To improve the sensitivity at the highest energies, HAWC is being upgraded with an outrigger array. It consists of 350 comparably much smaller water Cherenkov detectors, sparsely distributed around the HAWC main array. It will increase the instrumented area by a factor of 4-5. In this contribution, we will present the current status of the upgrade as well as simulation results on anticipated improvements in the performance of the observatory.}

\FullConference{35th International Cosmic Ray Conference --- ICRC2017\\
		10--20 July, 2017\\
		Bexco, Busan, Korea}

\begin{document}

\section{Introduction}

The gamma-ray sky above tens of TeV energies is still rarely studied and is of great interest. Sources emitting gamma-rays at those energies might be associated to PeVatrons that accelerate cosmic rays to PeV energies. Additionally, the study of diffuse emission or extended sources at these energies may shed light to more exotic phenomena. Therefore, it is prudent to have a detector able to perform a detailed survey of a large fraction of sky with enough sensitivity at the highest energies. The HAWC observatory \cite{ref_HAWC}, which has been fully operational since March 2015 is an excellent instrument to serve this purpose, detecting high energy cosmic and gamma rays from 2/3 of the celestial sphere every day.

\subsection{HAWC Observatory}

\begin{figure}[h!]
\centering
\includegraphics[width=.48\linewidth]{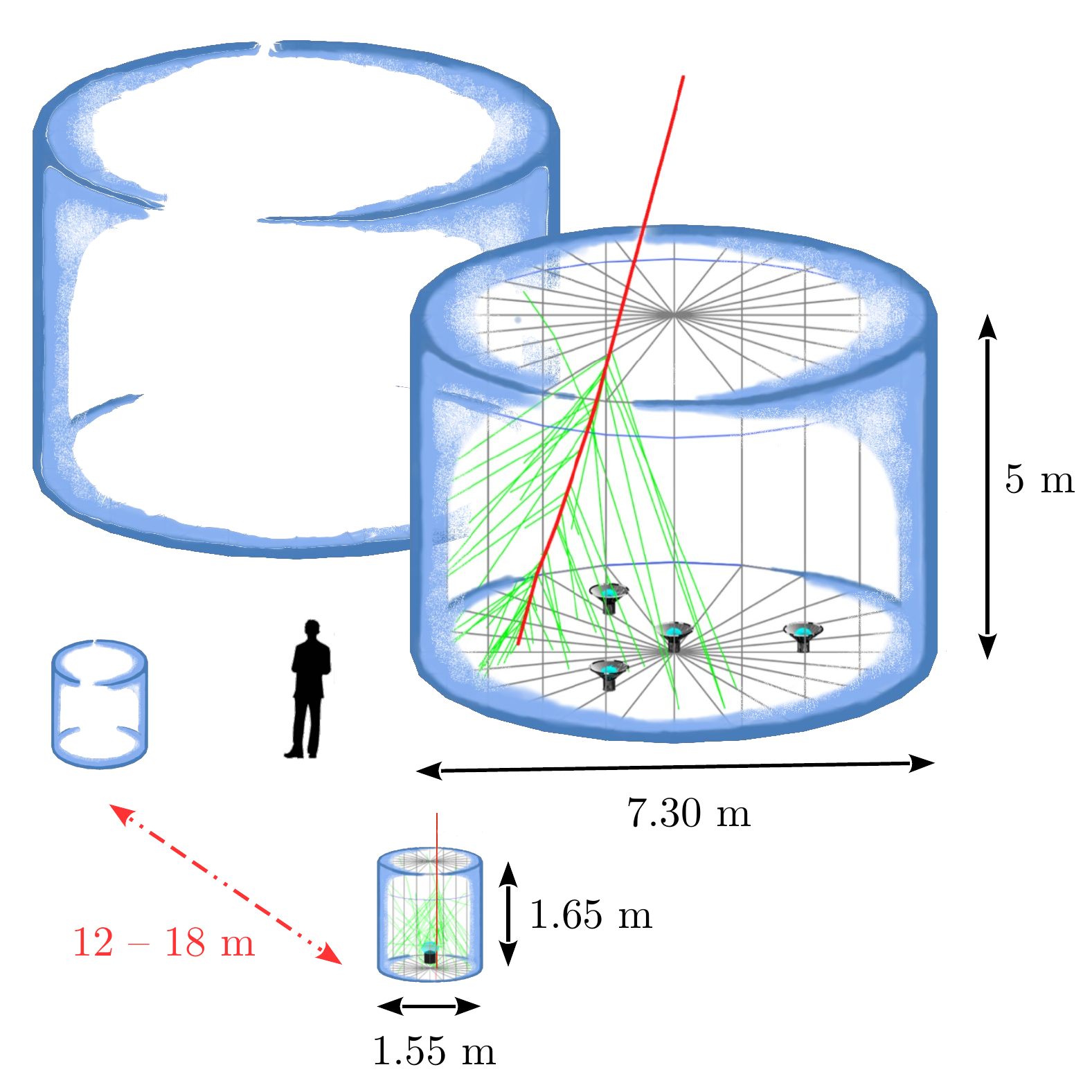}
\includegraphics[width=.48\linewidth]{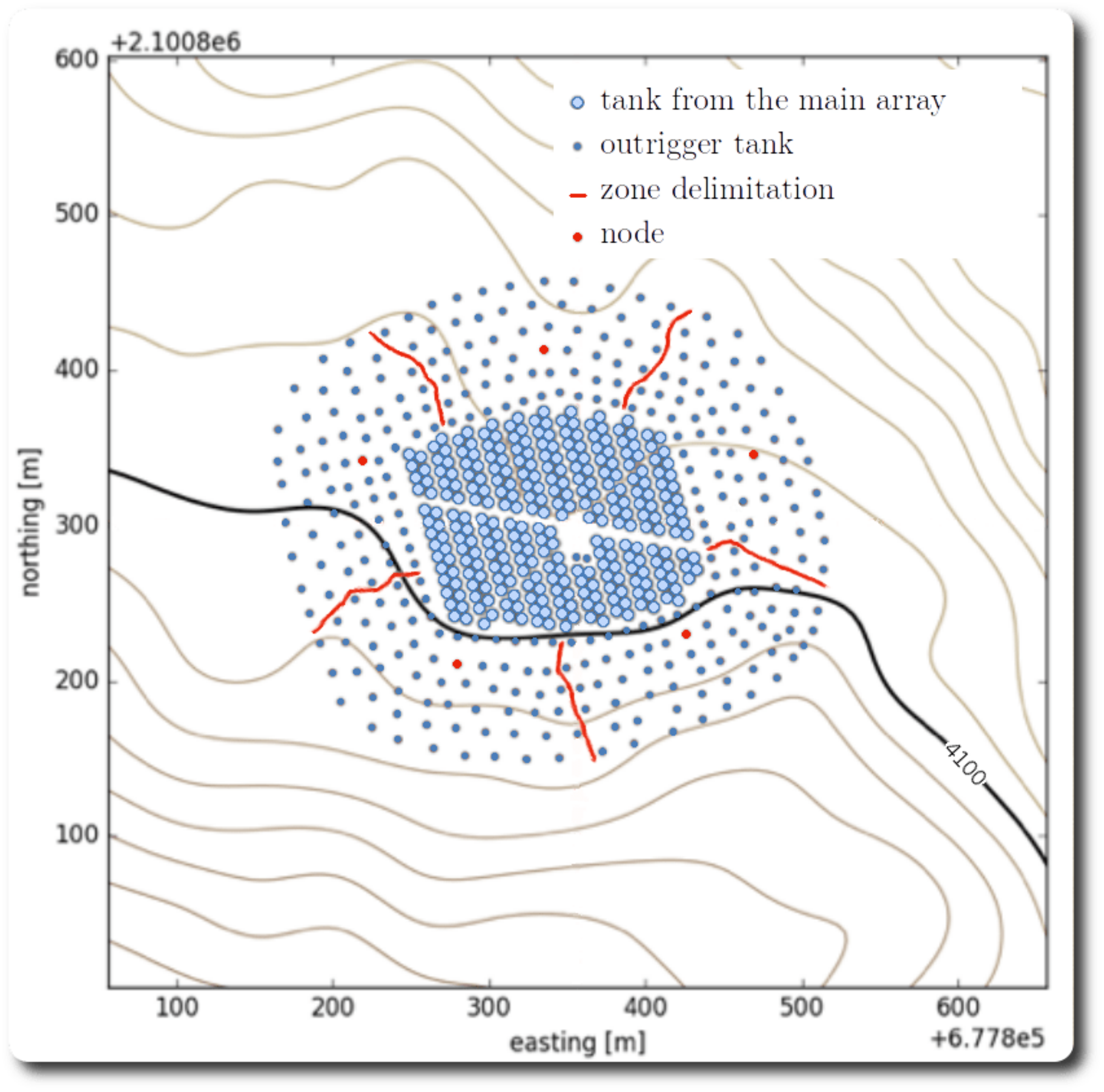}
\caption{Left panel: Schematics of main array (top right) and outrigger (bottom left) tanks. Right panel: Outrigger array surrounding the main HAWC array. The red lines shows the different sections of the outrigger array. The red dots show the node locations hosting the trigger and readout electronics.}
\label{Outrigger_array_layout_schematic}       
\end{figure}

The HAWC observatory  is a successor of the Milagro observatory \cite{ref_Milagro}, based on the water Cherenkov detection technique, which involves the detection of Cherenkov light produced in water by the secondary particles generated in an atmospheric air shower. It is situated on the flanks of the Sierra Negra peak in Central Mexico at an altitude of 4100 m above sea level. It consists of 300 Water Cherenkov Detectors (WCDs) in the main array encompassing a surface area of 22000 m$^2$. It has a wide field of view of 2 sr and an operational energy range of 0.1-100 TeV. The HAWC main  array WCDs are comprised of big cylindrical water tanks (see Figure \ref{Outrigger_array_layout_schematic} left panel)  of diameter 7.3 m and height 4.5 m equipped with four (three 8" and one 10") upward facing Photo Multiplier Tubes (PMTs) anchored at the bottom of the tank. 

\subsection{Motivation for the Upgrade}

The footprint of the shower on the ground is inherently dependent on the primary particle energy and on the altitude of the detector plane. At HAWC altitude the footprint of the shower at around tens of TeV energy of the primary particle becomes comparable to the total detector surface area. Therefore, most of the showers at these energies are not well-contained within the array. Although, the HAWC main array still has enough information in order to do gamma-hadron separation, direction reconstruction and shower size estimation but there is an ambiguity present because of large uncertainty in the core location. To tackle this situation the construction of an outrigger array \cite{ref_outriggers} around the main array has started. Now using the outrigger array, it will be possible to constrain the core location effectively enough so that these ambiguities in the shower reconstruction can be resolved. It will lead to increased number of well-reconstructed showers above multi-TeV energies. Hence, it  will improve the sensitivity of HAWC at those energies. One such outrigger array for Milagro instrument has already shown its effectiveness by dramatically increasing its sensitivity at the highest energies.

\section{Outrigger Array Description}

HAWC outrigger array will consists of 350 cylindrical tanks (see Figure \ref{Outrigger_array_layout_schematic}) of diameter 1.55 m and height 1.65 m with one 8" PMT \cite{ref_outriggers_2} anchored at the bottom of each tank. The outrigger array is being deployed in a concentric circular symmetric way around the main array covering an additional instrumented area of 4-5 times of the main array. The outriggers are mutually separated from each other by 12 to 18 m. The smaller size and larger separation of the outrigger WCDs are prompted by the fact that there are a lot of particles and consequently bigger signals present near to the core of a big shower. For trigger and readout purpose, the outrigger array is divided into 5 sections with 70 outriggers in each of them connected to a node (see Figure \ref{Outrigger_array_layout_schematic} right panel) with equal cable lengths. Each node will host the power supply and the trigger, readout and calibration system for the corresponding section.

\section{Trigger, Readout and Calibration for Outriggers}
\subsection{Trigger and Readout}
 The readout and trigger electronics is named as Flash Adc eLectronics for the Cherenkov Outrigger Node (FALCON). It will use the readout electronics developed for FlashCAM \cite{ref_flashcam}, one of the proposed cameras for medium-size telescopes for the Cherenkov Telescope Array (CTA). The motivation behind using the FlashCAM electronics in FALCON is the equivalence between each PMT of the outrigger array and a pixel of an Imaging Atmospheric Cherenkov Telescope (IACT) camera. One FALCON unit will accommodate three Flash-ADC boards, each of them can digitize 24 channels with a sampling speed of 250 MHz with a 12-bit accuracy. It also allows a flexible multiplicity trigger as well as the readout of full waveforms. The trace length is typically 40 samples (160 ns). The recorded wavelengths are used for charge extraction and signal timing information. 

Corresponding to each Flash-ADC board each outrigger section is divided in 3 sub-sections hosting maximum 24 outriggers. To trigger the outriggers, in any of the sub-section, at least 2 outrigger tanks should observe photo-electrons (pe) more than the defined threshold. If this condition is met the event will be readout for the outriggers.

\subsection{Photomultiplier tube calibration}

The PMTs used for the outrigger tanks are 8" Hamamatsu R5912. We performed a full calibration of these PMTs in a dark room, using a calibrated laser system of 398 nm wavelength. The single pe distribution was obtained using a very low intensity of the laser and the single pe calibration following the data driven approach described in \cite{ref_pmt_cal}.

\begin{figure}[h!]
\centering
\includegraphics[width=0.49\linewidth]{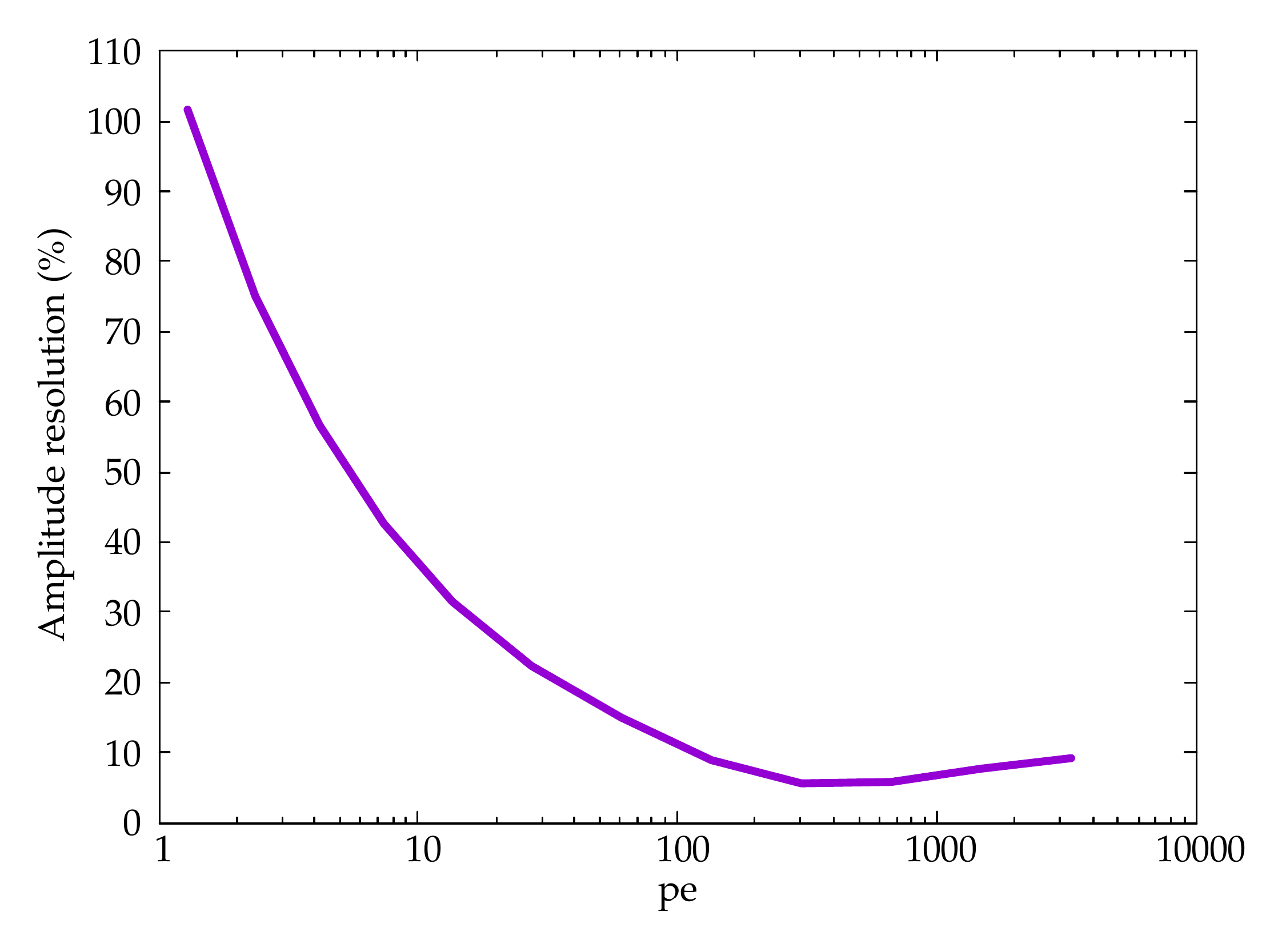}
\includegraphics[width=0.49\linewidth]{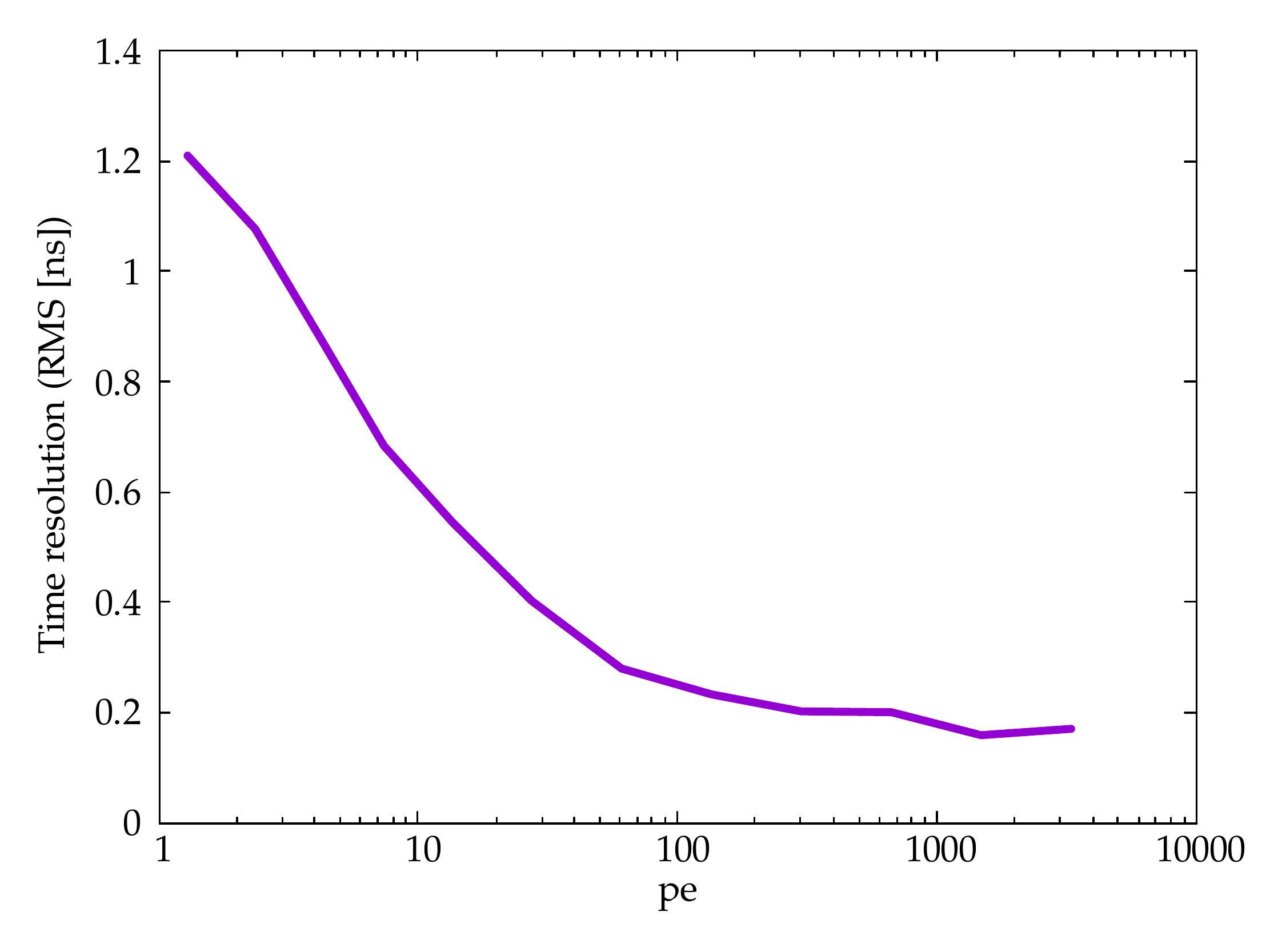}
\caption{Optimal calibration results for a  Hamamatsu 8" R5912 PMT, amplitude resolution (left), time resolution (right). Both curves show good enough resolutions for our purpose, even above 100 pe.}
\label{pmt_cal_plot}       
\end{figure}

To optimize the performances, in particular the amplitude and time resolution, a wide range of high voltage and laser intensity were tested, as well as different attenuation values to assess the trade-off between amplitude resolution and the dynamic range of the signal recorded by the FADC. The resulting amplitude and time resolution is shown on Figure \ref{pmt_cal_plot} which results in a 1500 V high voltage and an additional 6 db attenuation. We conclude that we have an amplitude resolution of 10\% and a time resolution of 200 ps above 100 pe. Since particle density fluctuations within a shower are typically larger than the resolution we get for both the time and the amplitude, this read out system is well suited for this application.

\subsection{First measurements on site}

The deployment of the outrigger array started with a first row of 10 tanks along the main array.
\begin{figure}[h!]
\centering
\includegraphics[width=0.85\linewidth]{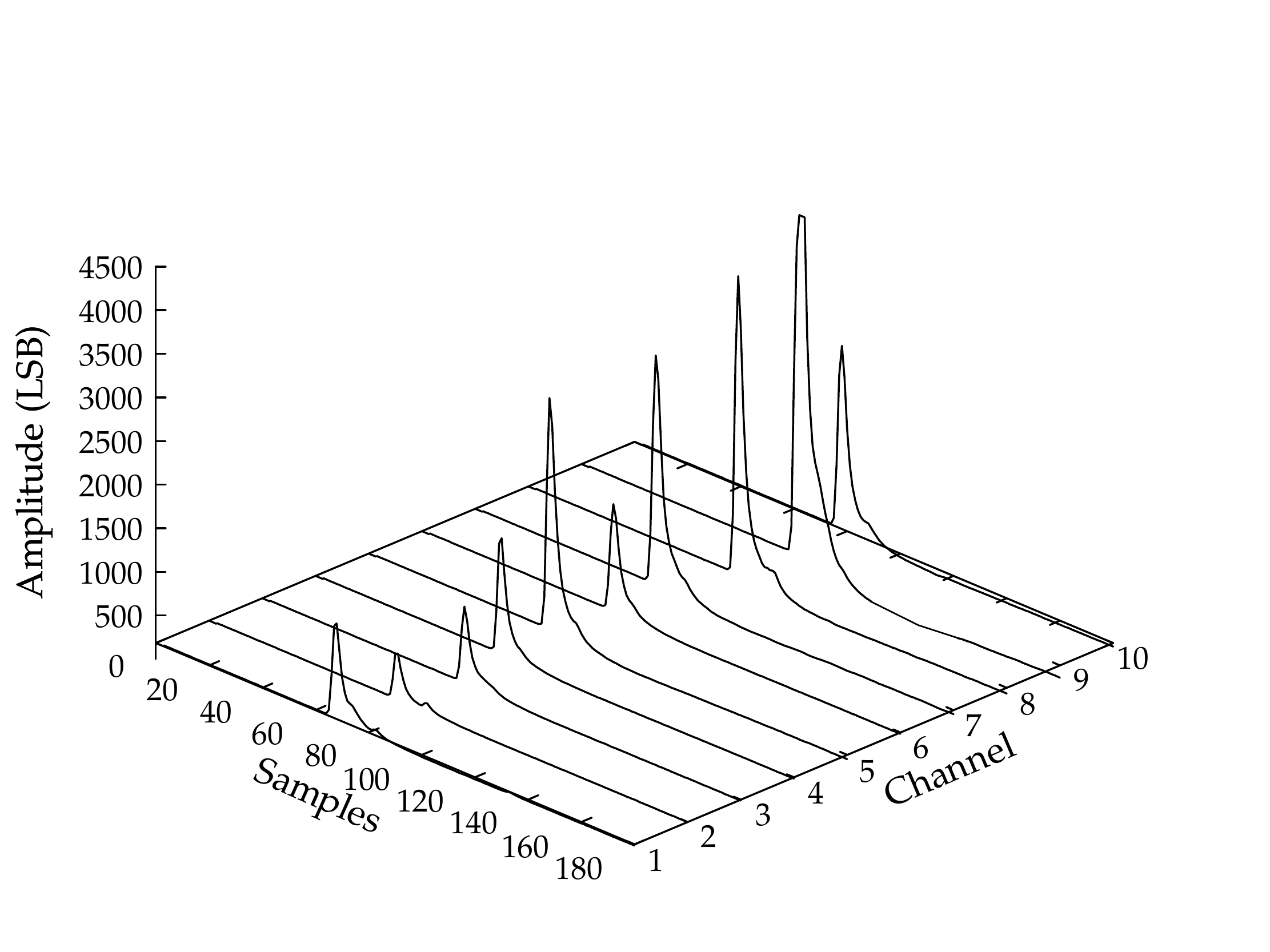}
\caption{Typical event with 10 outrigger tanks triggered within 160 ns. One sample is 4 ns.}
\label{event_outriggers}       
\end{figure}
They are currently operational and provide real data to test the FALCON electronics and study the trigger rate for different threshold values and different multiplicity trigger. 
This first set of tanks allowed us to confirm that we can run a 2-fold multiplicity trigger in a window of 200 ns. We measure a trigger rate of 700 Hz at a threshold of 1 pe, corresponding to the specifications for which it has been designed. The final threshold settings have not been optimized yet, therefore the final trigger rate of the system is subject to vary. Figure \ref{event_outriggers} shows a typical event where 10 tanks triggered.

\section{Simulation Results}
In this section, we present the results of the improvement of the reconstruction and detection of air showers obtained using the outrigger array, especially for the air showers that are falling outside the main array.

\subsection{Reconstruction Using Outriggers}

\begin{figure}[h!]
\centering
\includegraphics[width=0.85\linewidth]{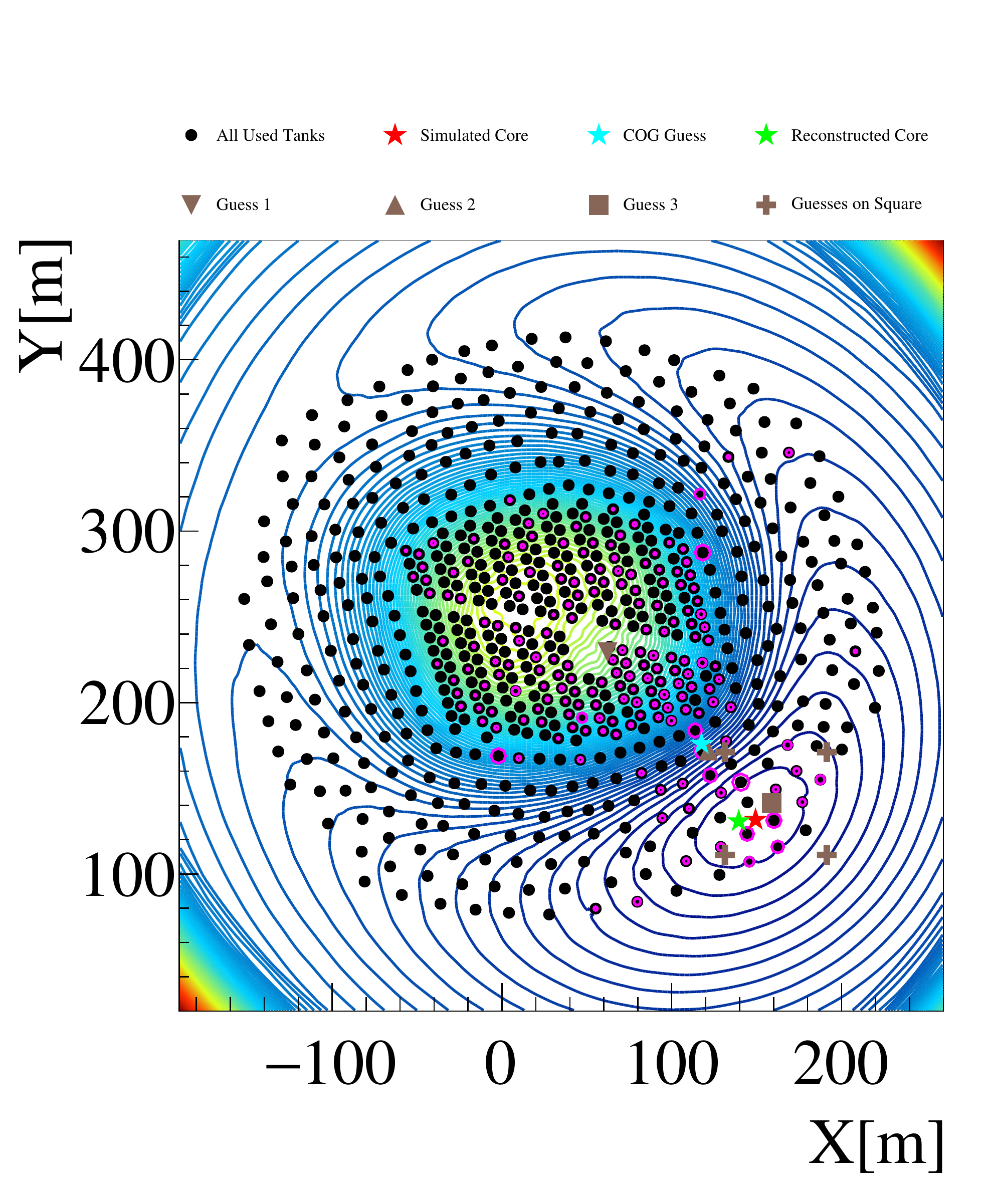}
\caption{Typical example of shower reconstruction combining HAWC main array and outriggers, using the likelihood fit method. The contours represents the 2D projection of the likelihood surface. The likelihood surface shows the value of likelihood as a function of location while other parameters are the best fit values. The blue colour of the contours shows the likelihood surface minimum and the red ones show the maximum. The magenta colour circles over the tanks represents the charge observed. The size of the circle is charge dependent.}
\label{likelihood_surface_HAWC+OR}       
\end{figure}

For the reconstruction of the air showers combining the main array and outriggers we have been developing a new reconstruction method. The method consists of a fit with a maximum likelihood approach, of the lateral amplitude distribution of the observed charge for a given shower on the detector plane against a Monte Carlo template based fit model. This method can be used to estimate the core location of the showers falling on the main array as well as on the outriggers. In addition to fitting the core, this method also estimates the primary particle's energy and the depth of the shower maximum (X$_{max}$). 

\begin{figure}[h!]
\centering
\includegraphics[width=0.85\linewidth]{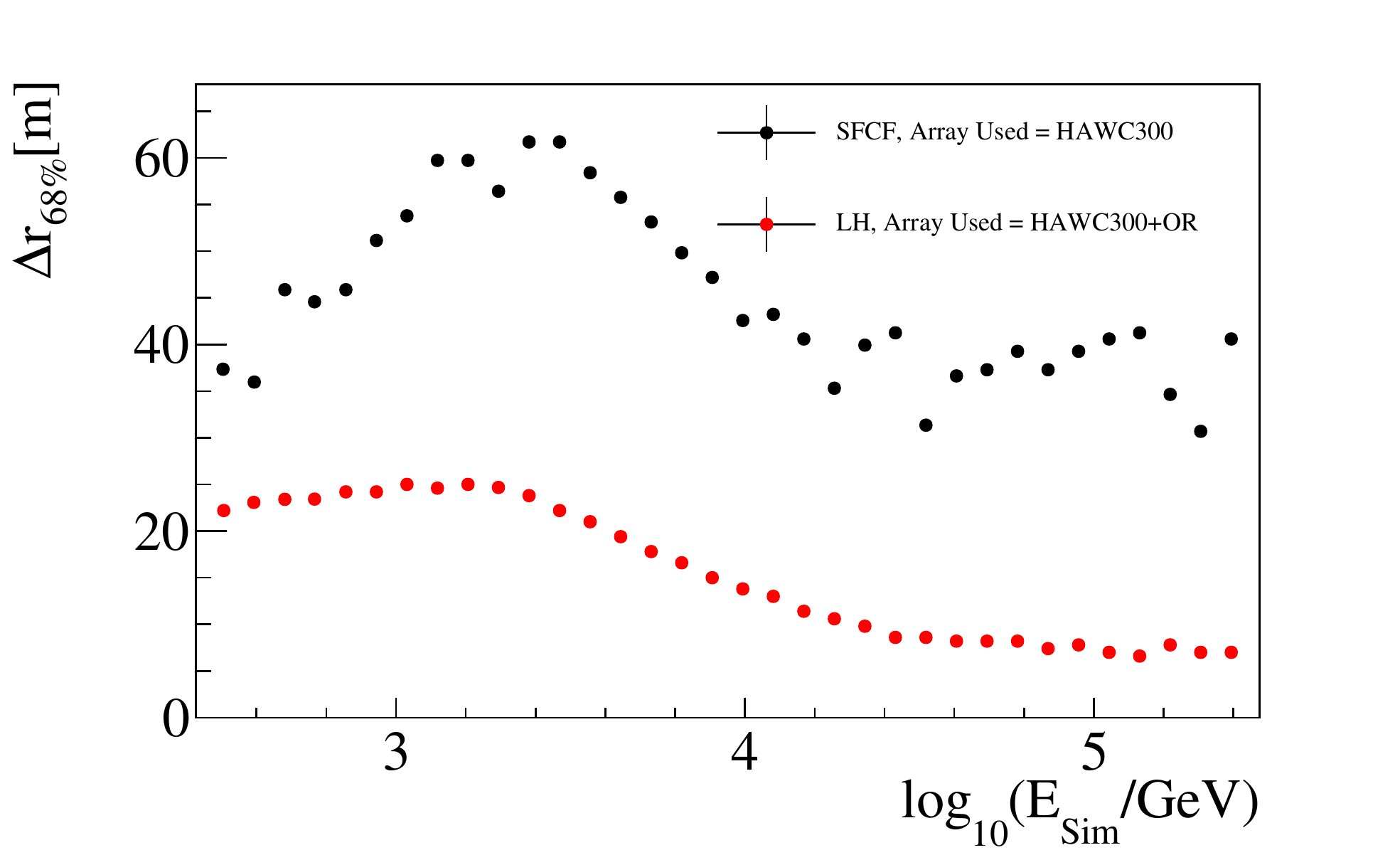}
\caption{Core resolution obtained vs. simulated energy in log$_{10}$ scale (log$_{10}$(E$_{Sim}$/GeV)). The points in each energy bin represent the 68\% containment area of the core$_{Simulated}$-core$_{Reconstructed}$ distribution. SFCF is the present core fit method being used in HAWC and LH is the likelihood method presented here. Array used means the array used for the reconstruction. All showers were thrown on the outriggers as shown in Figure \ref{Outrigger_array_layout_schematic}.}
\label{Core_resolution_comparison}       
\end{figure}

Figure \ref{likelihood_surface_HAWC+OR} shows a typical example of event reconstruction. We start with multiple initial core guesses as shown in the figure with brown colour markers. The location of these guesses is dependent on the center of mass (COM) guess shown as sky blue star. The motivation behind giving multiple guesses is  to avoid the possibility to get stuck in a local minimum of a multi-dimensional likelihood surface. The COM guess was calculated using the charge observed in the different channels shown with magenta coloured circles over the tanks.  The simulated and reconstructed cores are shown as red and green stars respectively. 

For the evaluation of the improvement in the core resolution by using the outriggers, we simulated gamma-ray events on the outrigger array only. We did the reconstruction in two different scenarios: 

\begin{itemize}
\item[1.] Using the core fitter presently being used in HAWC known as super fast core fitter (SFCF) \cite{ref_SFCF} on the HAWC main array.

\item[2.] Using the likelihood (LH) fit method on the HAWC main array combining with outriggers.
\end{itemize}

The maximum zenith angle reconstructed was 45$^\circ$. We also put a threshold of 20 tanks as the minimum number of tanks hit with at least a single pe from the main array. Figure \ref{Core_resolution_comparison} depicts that using the new likelihood fit and adding the outriggers gives us an improvement of $\sim$75\% on the core resolution above a few TeV energies. The improvement in the core resolution is very promising and certainly, it will ameliorate the reconstruction of the showers falling outside the main array.     

\section{Current Status and Outlook}

The deployment of the outrigger array is currently taking place. The first few outriggers are already taking data on site using the FALCON electronics. The FALCON readout integration to the central DAQ will be finished soon. The software for the reconstruction chain is already in a very good shape and the first results using simulations look encouraging. The next step is to deploy the sections one by one, the deployment of the first one being imminent. The full outrigger array is planned to be completed by the beginning of next year. 

A fully functional outrigger array will be able to  resolve the uncertainties in determining the core location for big showers falling outside the main array. That will lessen the ambiguities in the shower reconstruction and hence will improve the sensitivity at the highest energies. 

\section*{Acknowledgements}

We acknowledge the support from: the US National Science	Foundation (NSF); the US Department	of Energy Office of High-Energy	Physics; the Laboratory	Directed Research and	Development (LDRD) program of Los Alamos National Laboratory;	Consejo	Nacional de	Ciencia	y Tecnolog\'{\i}a	(CONACyT), M{\'e}xico (grants 271051, 232656, 260378, 179588,	239762,	254964,	271737,	258865,	243290,	132197),	Laboratorio	Nacional HAWC de rayos gamma; L'OREAL Fellowship for Women in	Science	2014; Red HAWC,	M{\'e}xico;	DGAPA-UNAM (grants RG100414, IN111315, IN111716-3, IA102715, 109916, IA102917); VIEP-BUAP;	PIFI 2012, 2013,	PROFOCIE 2014, 2015; the University	of Wisconsin Alumni	Research Foundation; the Institute of Geophysics, Planetary	Physics, and Signatures	at Los Alamos National Laboratory;	Polish Science Centre grant DEC-2014/13/B/ST9/945; Coordinaci{\'o}n de	la Investigaci{\'o}n Cient\'{\i}fica	de la Universidad Michoacana. Thanks to Luciano	D\'{\i}az	and	Eduardo	Murrieta for technical support.


\begin{thebibliography}{99}
\bibitem{ref_HAWC}
HAWC Collaboration, A.U. Abeysekara et al., Astropart. Phys. 50-52 (2013) p26-32.
\bibitem{ref_Milagro}
Milagro Collaboration, R.W. Atkins et al., ApJ 595 (2003) p801-811.
\bibitem{ref_outriggers}
A. Sandoval, Proc. of the 34rd ICRC, The Hague, The Netherlands, September (2015), astro-ph.IM:1509.04269.
\bibitem{ref_outriggers_2}
V. Joshi, Proc. RICAP 2016, astro-ph.IM:1701.06376
\bibitem{ref_flashcam}
G. P\"{u}hlhofer. C. Bauer, F. Eisenkolb, D. Florin, C. F\"{o}hr, A. Gadola, G. Hermann, C. Kalkuhl, J. Kasperek, T. Kihm, J. Koziol, A. Manalaysay, A. Marszalek, P. J. Rajda, W. Romaszkan, M. Rupinski, T. Schanz, S. Steiner, U. Straumann, C. Tenzer, A. Vollhardt, Q. Weitzel, K. Winiarski, K. Zietara, and f. t. CTA consortium, Proc. of the 33rd ICRC, Rio de Janeiro, Brazil, July (2013), astro-ph.IM:1307.3677 .
\bibitem{ref_pmt_cal} 
R. Saldanha, L. Grandi, Y. Guardincerri, T. Wester, Model Independent Approach to the Single Photoelectron Calibration of Photomultiplier Tubes, astro-ph.IM, 2016
\bibitem{ref_SFCF} HAWC Collaboration, A.U. Abeysekara et al., Observation of the Crab Nebula with the HAWC Gamma-Ray Observatory,astro-ph.HE:1701.01778 
\end{thebibliography}
\end{document}